\documentclass[11pt,a4paper]{article}
\usepackage[utf8]{inputenc}
\usepackage{amsmath}
\usepackage{mathtools}
\usepackage{a4wide}
\usepackage{cite}
\usepackage{authblk}
\usepackage{graphicx}
\usepackage{amsmath,amssymb}
\usepackage[enable]{easy-todo}
\usepackage[hang,small,bf]{caption}
\usepackage[subrefformat=parens]{subcaption}

\title{Lower bounds on lepton flavor violating branching ratios \\in a radiative seesaw model}
\author{Osamu Seto$^{a,b}$, Tetsuo Shindou$^{c}$, Takanao Tsuyuki$^{c,d}$\footnote{tsuyuki@cc.kogakuin.ac.jp}}
\affil{\small $^a$\textit{Institute for the Advancement of Higher Education, Hokkaido University, Sapporo 060-0817, Japan}}
\affil{\small $^b$\textit{Department of Physics, Hokkaido University, Sapporo 060-0810, Japan}}
\affil{\small $^c$\textit{Division of Liberal-Arts, Kogakuin University, Hachioji, Tokyo 192-0015, Japan}} 
\affil{\small $^{d}$\textit{Academic Support Center, Kogakuin University, Hachioji, Tokyo 192-0015, Japan}}
\date{}
\begin{document}

\maketitle
\begin{abstract}
We study the lepton flavor violating decays such as $\mu\to e\gamma$, $\tau\to e\gamma$, 
$\tau\to \mu\gamma$ in the three-loop radiative seesaw model proposed by Krauss, Nasri, and Trodden. 
In this model, the relevant coupling constants are larger for the heavier scalars that run inside loop diagrams to generate the appropriate magnitude of neutrino masses.
Imposing a criterion that all the coupling constants must be small enough to be 
treated perturbatively, we find an upper bound on the mass of one of the scalars. 
By combining it with neutrino mass parameters, we derive lower bounds on the branching ratios of the lepton flavor violating processes. 
In a case with the inverted mass ordering and best-fit neutrino oscillation parameters, one of the lower bounds
is $\text{Br}(\mu\to e\gamma)>1.1\times 10^{-13}$, which 
is within the reach of the MEG~II experiment.
\end{abstract} 

\begin{flushleft}
EPHOU-22-002, KU-PH-031
\end{flushleft}

\section{Introduction}
The origin of the neutrino masses has been a big mystery in the Standard Model (SM) of particle physics and indicates new physics. 
Several mechanisms are proposed to explain the tininess of the neutrino masses in the literature. For example, the seesaw model is a popular one in that the enormous Majorana neutrino mass scale suppresses the mass scale of the light neutrinos~\cite{Minkowski:1977sc,Yanagida:1979as,GellMann:1980vs,Mohapatra:1979ia}. The Majorana mass is naturally required to be larger than $10^6$~GeV. 

Utilizing the higher-loop suppression is an alternative approach: the neutrino mass matrix is radiatively generated in models along this line. A. Zee proposed the first concrete model in 1980\cite{Zee:1980ai}, and many models have been proposed since then\cite{Cheng:1980qt,Zee:1985id,Babu:1988ki,Krauss:2002px,Ma:2006km, Aoki:2008av}. A comprehensive review is provided, for example, in Ref.\cite{Cai:2017jrq}.
Since the new particles are relatively light
and have large couplings with the SM particles,
those models are testable by experiments.

There is a class of models, the so-called radiative seesaw models, where right-handed neutrinos are introduced\cite{Krauss:2002px,Ma:2006km, Aoki:2008av}. 
To avoid their tree-level contribution to the neutrino masses, additional $Z_2$ symmetry will be necessary, and the right-handed neutrinos have an odd charge.
The $Z_2$ symmetry simultaneously guarantees the stability of the lightest right-handed neutrino, and it can be dark matter.

The Krauss-Nasri-Trodden (KNT) model is an example of the radiative seesaw model, where the neutrino masses are generated via the three-loop diagrams \cite{Krauss:2002px}. 
This model includes right-handed neutrinos, new scalar particles $S_1$ and $S_2$.
Because of the three-loop suppression, the coupling constants in this model tend to be as large as $\mathcal{O}(1)$ if the new particles are as heavy as TeV. 
The perturbative treatment breaks down when the coupling constants are much larger than $\mathcal{O}(1)$. In order to avoid this, the coupling constants are preferred to be smaller than one. 
It indicates that the new particle mass scale will have an upper bound.
On the other hand, the new particles cause lepton flavor violating (LFV) processes such as   $\mu\to e\gamma$, $\tau\to e\gamma$, and $\tau\to \mu\gamma$\cite{Ahriche:2013zwa,Chowdhury:2018nhd,Cepedello:2020lul}, and it is more significant for the lighter new particles. 

In this paper, we study the details of these constraints in the KNT model. 
We first show the feature of the loop function, and show that it provides the upper bound on the mass of $S_1$. 
Then we study the lower limit of the predicted LFV processes.
In the KNT model, the Yukawa matrix with $S_1$ has an asymmetric flavor structure, 
and we can use a valuable technique provided in Ref.~\cite{Irie:2021obo} to determine the parameters in the Yukawa matrices.
We derive the lower bounds on the branching ratios of LFV decays.
For the case with the three right-handed neutrinos and the inverted mass ordering, the lower bounds are $\text{Br}(\mu\to e\gamma)>1.1\times 10^{-13}$, $\text{Br}(\tau\to e\gamma)>1.5\times 10^{-14}$, and $\text{Br}(\tau\to \mu\gamma)>3.7\times 10^{-13}$ when neutrino oscillation parameters are best-fit values. Interestingly, the lower bound of $\text{Br}(\mu\to e\gamma)$ is within the reach of the MEG II experiment \cite{MEGII:2021fah}.

This paper is organized as follows. 
In Sec.~\ref{snumass}, we briefly introduce the model, and 
we describe the neutrino mass matrix. 
In Sec.~\ref{slo}, we discuss the upper bound on the mass of $S_1$.
In Sec.~\ref{sbr}, we derive the lower bound of the LFV branching ratios
induced by the $S_1$ exchange. 
In Sec.~\ref{secBP}, we consider a benchmark scenario and 
discuss the $S_2$ contribution to $\text{Br}(\tau\to\mu\gamma)$.
We give a conclusion in Sec.~\ref{scon}.

\section{The KNT model} \label{snumass}
We consider the KNT model\cite{Krauss:2002px}.
Two charged scalars $S_1$, $S_2$ and $n_N$ right-handed neutrinos $N_I$ ($I=1,\ldots n_N$) are introduced.
It is known that $n_N\geq 2$ is necessary to reproduce the neutrino mass matrix
consistent with the neutrino oscillation experiments\cite{Cheung:2004xm}.
The discrete $Z_2$ symmetry
\begin{align}
Z_2: \{S_2, N_I\} \to \{-S_2, -N_I\}
\end{align}
is imposed to forbid the Dirac masses of neutrinos.
This symmetry also guarantees the stability of the lightest $Z_2$ odd particle, 
which is a dark matter candidate. 
Therefore, the lightest $Z_2$ odd particle should be electrically neutral, 
and its relic abundance in the early Universe should be less than 
the observed dark matter relic abundance $\Omega h^2\simeq 0.1$.
The standard model Lagrangian is extended by the terms
\begin{align}
  \mathcal{L}_{\text{KNT}}=
\frac{h_{ij}}{2}\overline{L_i^c}i\tau_2 L_j S_1^+ 
+ g_{Ij}^*\overline{N_I^c}\ell_{Rj}S_2^+
+\frac{m_{N_I}}{2}\overline{N_{I}^c}N_{I}+\textrm{h.c.}-V.
\label{lagKNT}
\end{align}
We choose the flavor basis such that both the charged lepton mass matrix and the right-handed neutrino mass matrix are diagonal with real and positive elements. 
The lepton doublet is $L=
(\nu_{L},\ \ell_{L})$, and the convention of the neutrino mixing matrix is
\begin{equation}
U=
\begin{pmatrix}
	c_{12}c_{13}&s_{12}c_{13}&s_{13}e^{-i\delta}\\
	-s_{12}c_{23}-c_{12}s_{13}s_{23}e^{i\delta}&
	c_{12}c_{23}-s_{12}s_{13}s_{23}e^{i\delta}&
	c_{13}s_{23}\\
	s_{12}s_{23}-c_{12}s_{13}c_{23}e^{i\delta}&
	-c_{12}s_{23}-s_{12}s_{13}c_{23}e^{i\delta}&
	c_{13}c_{23}
\end{pmatrix}
\begin{pmatrix}
e^{i\eta}&0&0\\
0&1&0\\
0&0&e^{i\eta'}	
\end{pmatrix}\;.
\end{equation}

\begin{figure}
\begin{center}
\includegraphics{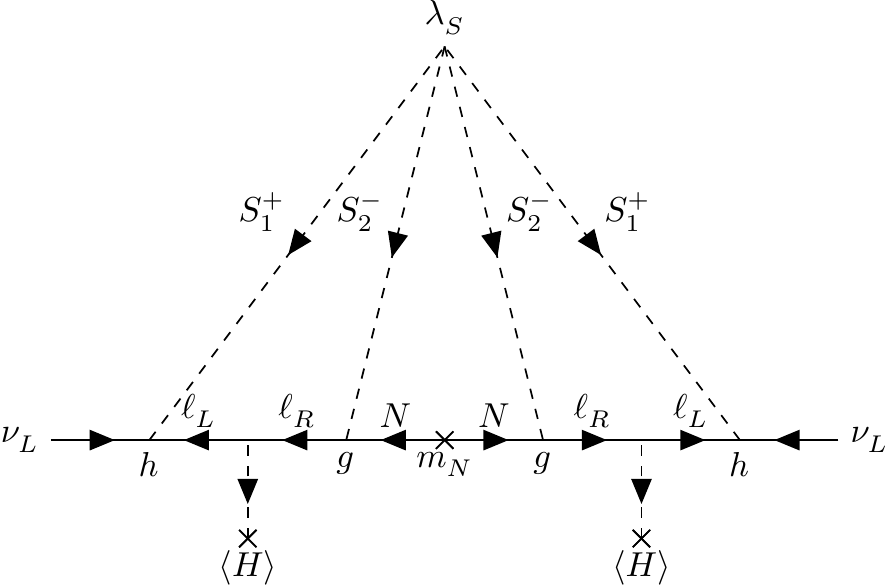}
\end{center}
\caption{The diagram relevant to the neutrino mass matrix in the KNT model.\label{Fig:MnuInKNTmodel}}
\end{figure}

In this model, the neutrino mass matrix arises from the three-loop diagram shown in Fig.~\ref{Fig:MnuInKNTmodel},
and each element is given by\cite{Ahriche:2013zwa}
\begin{align}
M_{ab} &=
\frac{\lambda_{S}}{4(4\pi)^3m_{S_1}}\sum_{I,j,k}m_{\ell_j}m_{\ell_k}	h_{aj}h_{bk}g_{Ij}
g_{Ik}f(x_I,y), \label{eq:KNTneutrinomassmatrix}\\
x_I &\equiv \frac{m_{N_I}^2}{m_{S_2}^2},\ y \equiv \frac{m_{S_1}^2}{m_{S_2}^2},
\end{align}
where $f(x,y)$ is the loop function, and 
$\lambda_S$ is a coupling constant of the scalar quartic coupling 
included in the potential $V$ as 
\begin{align}
V&\supset \frac{\lambda_S}{4}(S_1^-)^2(S_2^+)^2+\text{H.c.}\ .
\end{align}
\begin{figure}[tb]
\centering
\includegraphics[width=90mm]{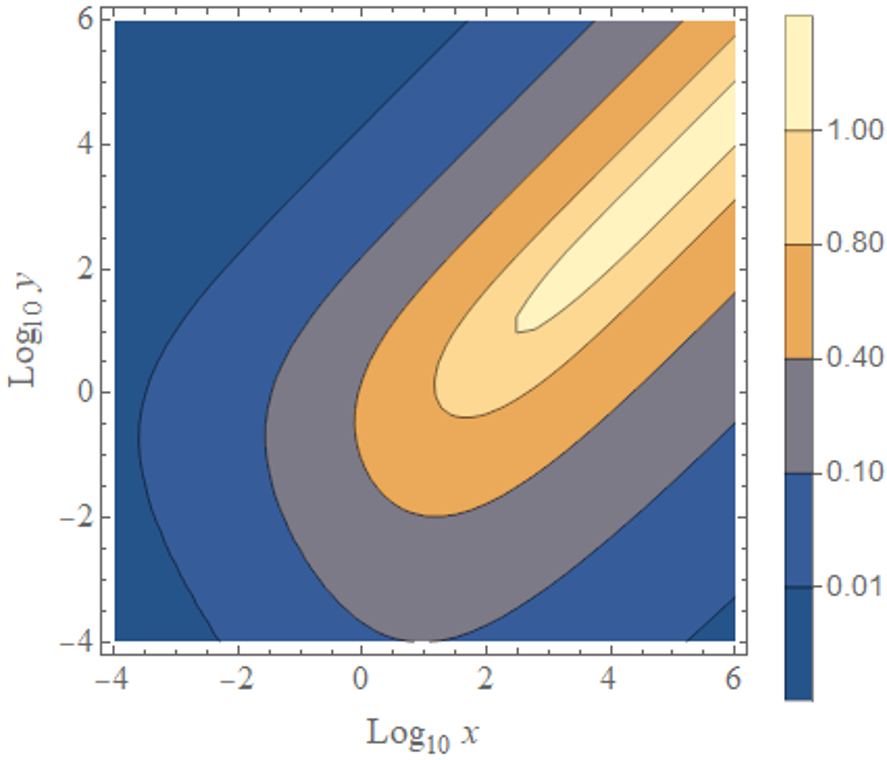}
\caption{Contour plot of $f(x,y)$. It has maximal value 1.044 when $x/y=10^{1.47}$, $y\gg 1$.}  \label{fbf1}
\end{figure}

\begin{figure}[tb]
\centering
\includegraphics[width=90mm]{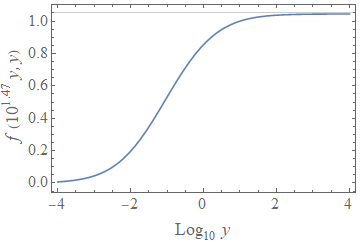}
\caption{Loop function $f(x,y)$ with $x/y=10^{1.47}$. The horizontal line shows the upper bound 1.05. The loop function becomes constant for large $y$ [see Eq. (\ref{efap})].} \label{fbf3}
\end{figure}

Here, we discuss the behavior of the loop function $f(x, y)$ and show that there is 
an upper bound on $f(x, y)$. The loop function is calculated as\footnote{There is an additional term $\ln[r(\eta_+-1)(1-\eta_-)]$ in the definition of $I(r, y)=-J(r,y)/r$ in Ref. \cite{Chowdhury:2018nhd}, where $\eta_{+}=(q-y)/r$, $\eta_{-}=(1/q-1)y/r$. We confirmed that this term is identically zero.}
\begin{align}
    f(x, y)&=\frac{\sqrt{x}}{8y^{3/2}}\int_0^\infty dr \frac{J(r,y)^2}{r(r+x)},\\
	J(r,y)&=q\ln\left[\frac{y}{q}\right]+\frac{y}{q}\ln[q]+(1+r)\ln\left[\frac{1+r}{y}\right],\\
	q &= \frac{1}{2}\left(1+r+y+\sqrt{(1+r+y)^2-4y}\right).
\end{align}
Note that our definition of the loop function is different 
from the one in Ref.~\cite{Ahriche:2013zwa}.
The function $f$ is related to the function $F$ in Ref.~\cite{Ahriche:2013zwa}
as $f=\frac{m_{S_1}}{m_{S_2}}F$.
This redefinition is convenient to discuss the mass bound on the new particles. 
As shown in Figs.~\ref{fbf1} and \ref{fbf3}, the value of $f$ is saturated.
On the other hand, $F$ is not bounded from above.

This property of $f$ can be analytically understood as follows. 
First, we consider the $x$ dependence of $f(x,y)$ for a fixed value of $y$.
The asymptotic behavior of $f$ is given by 
\begin{align}
f(x, y) &\propto \sqrt{x}\ (x\ll 1)\\
f(x, y) &\propto \frac{1}{\sqrt{x}}\ (x\gg 1, x\gg y)\;,
\end{align}
so $\displaystyle \lim_{x\to 0}f(x,y)=0$ and $\displaystyle\lim_{x\to \infty}f(x,y)=0$.
Thus $f$ has a maximum when $y$ is fixed.
Second, we consider the $y$ dependence for a fixed $x$.
For $y\ll 1$,
\begin{align}
f(x, y) &\propto \sqrt{y}(\ln y)^2 \to 0\ (y\to  0),
\end{align}
is satisfied, while one can find $F(x,y)\propto (\ln y)^2$ which diverges in the limit of $y\to 0$. 
For $y\gg 1$,
we find 
\begin{align}
f(x,y)&= \frac{1}{8}\sqrt{\frac{x}{y}} \int_0^\infty dt \frac{\left[(t+1)\ln(t+1)-t\ln t \right]^2}{t(t+x/y)}, \label{efap}
\end{align}
where we put $t\equiv r/y$. We can see that $f$ depends only on $x/y$ when $y\gg 1$. 
Focusing on the case of $x\ll y$, we find
\begin{align}
f(x,y)&\propto \sqrt{\frac{x}{y}} \to 0 \ (y \to \infty).
\end{align}
Thus $f$ has a maximum when $x$ is fixed.

From the above discussion, we expect that $f(x,y)$ is maximized for $x\sim y\gg 1$. 
As shown in Eq.~(\ref{efap}), $f(x,y)$ depends only on $x/y$ for $y\gg 1$.
Using numerical evaluation we find that $f(x,y)$ is saturated to 1.044 as shown in Fig.~\ref{fbf3} in the direction of  $x/y\simeq 10^{1.47}$.
As a conclusion, we obtain the upper bound of $f(x,y)$ as 
\begin{align}
f(x,y) < 1.05. \label{ebfu}
\end{align}

\section{Upper bound on $S_1$ mass}\label{slo}

Let us adopt a criterion that all the dimensionless coupling constants in the Lagrangian Eq.~(\ref{lagKNT}) are less than unity.
If a coupling constant is larger than $\mathcal{O}(1)$, it blows up quickly by renormalization group running and
then the perturbative treatment is broken down. 
To avoid that we take into account the ansatz, and we show that it leads to the upper bound on the mass scale of 
new particles.

In Eq.~(\ref{eq:KNTneutrinomassmatrix}) we can naturally expect that the terms proportional to $m_e$ get strong suppression, and 
we can ignore such terms. In this approximation, three components of the neutrino mass matrix are given by
\begin{align}
M_{\mu\mu} &= \frac{\lambda_S m_\tau^2 h_{23}^2 }{4(4\pi)^3 m_{S_1}}\sum_{I=1}^{n_N} g_{I3}^{2} f(x_I,y)\;, \label{emmumu}\\
M_{\mu\tau} &= -\frac{\lambda_S m_\mu m_\tau h_{23}^2 }{4(4\pi)^3 m_{S_1}}\sum_{I=1}^{n_N} g_{I2}g_{I3} f(x_I,y)\;, \label{emmutau}\\
M_{\tau\tau} &= \frac{\lambda_S m_\mu^2 h_{23}^2 }{4(4\pi)^3 m_{S_1}}\sum_{I=1}^{n_N} g_{I2}^{2} f(x_I,y)\;. \label{etautau}
\end{align}
The observed neutrino oscillation data need that these components are the same order, in spite of $M_{\mu\mu}\propto m_{\tau}^2$, $M_{\tau\tau}\propto m_{\mu}^2$. 
It requires that some of the Yukawa couplings $g_{I2}$ are much larger than the other elements $g_{I3}$, so we impose the perturbativity condition to $g_{I2}$.

By using the triangle inequality, the ansatz $|g_{I2}|< 1$, and the upper bound of $f(x_I,y)$ [Eq. (\ref{ebfu})], we find
\begin{align}
\left| \sum_{I} g_{I2}^{2} f(x_I,y)\right| &\leq \sum_{I} |g_{I2}|^2 f(x_I,y) 
< \sum_{I} f(x_I,y)\nonumber\\
&<1.05n_{\text{eff}}\;,
\label{uppersum}
\end{align}
where $n_{\text{eff}}$ denotes the number of right-handed neutrinos that contribute to the summation. If $N_1$ is dark matter, we need $m_{N_1}<m_{S_2}$ (i.e., $x_1<1$), then $f(x_1,y) \ll 1.05\ (y\gg 1)$ as shown in Fig.~\ref{fbf1}.
Thus $g_{12}^2f(x_1,y)$ is negligible in $M_{\tau\tau}$ and
\begin{align}
n_{\text{eff}}\leq n_N-1
\end{align} 
is satisfied.
By using the inequality~(\ref{uppersum})  and $\lambda_S < 1$, we obtain
a upper bound on $m_{S_1}$ as 
\begin{align}
m_{S_1} &< \frac{m_\mu^2 |h_{23}|^2 }{4(4\pi)^3 |M_{\tau\tau}|}1.05n_{\text{eff}} \nonumber \\
&=7.39\times 10^4\ {\rm GeV} \left(\frac{\rm 0.02 eV}{|M_{\tau\tau}|}\right) |h_{23}|^2 n_{\text{eff}}\;.\label{ems1_2}
\end{align}
Similar upper bounds can be obtained by using the other components $M_{\mu\tau}$ and $M_{\mu\mu}$:
\begin{align}
m_{S_1} &<7.39\times 10^4\ {\rm GeV} \left(\frac{\rm 0.02 eV}{|M_{\mu\tau}|}\right) |h_{23}|^2 n_{\text{eff}}'\left(\frac{g_{I3,\text{max}}}{m_\mu/m_\tau}\right)\;,\label{ems1_mutau} \\
m_{S_1} &<7.39\times 10^4\ {\rm GeV} \left(\frac{\rm 0.02 eV}{|M_{\mu\mu}|}\right) |h_{23}|^2 n_{\text{eff}}''\left(\frac{g_{I3,\text{max}}}{m_\mu/m_\tau}\right)^2\;,\label{ems1_mumu}
\end{align}
where $n_\text{eff}'$ and $\ n_\text{eff}'' $ are the number of the right-handed neutrinos that contribute the corresponding neutrino mass components.
These bounds depend on the maximal value of $|g_{I3}|$ (denoted by $g_{I3,\text{max}}$), so we focus on the bound (\ref{ems1_2}). Since $g_{I3,\text{max}}$ is expected to be the order of $m_\mu/m_\tau$ as indicated in Eqs.~(\ref{emmumu})-(\ref{etautau}) [and also the example (\ref{egmat})], these bounds are the same order as the inequality (\ref{ems1_2}). The bounds (\ref{ems1_mutau}) and (\ref{ems1_mumu}) can be significant when $M_{\tau\tau}$ 
is suppressed.

The size of $M_{\tau\tau}$ is determined by the light neutrino masses and the mixing matrix elements as 
\begin{align}
M_{\tau\tau}=m_1 U_{3 1}^2+m_2 U_{3 2}^2+m_3 U_{3 3}^2\;.
\end{align}
In the KNT model, the Yukawa matrix $h_{ij}$ is antisymmetric, so
the determinant of the neutrino mass matrix vanishes. 
It means that $m_1=0$ for the normal ordering (NO) case and $m_3=0$ for the inverted ordering (IO) case. 
Since the Majorana phases have not been restricted by the experiments, the range of $|M_{\tau\tau}|$ 
is determined by the triangle inequality:
\begin{align}
\left|m_2|U_{32}|^2-m_3|U_{33}|^2\right| &\leq |M_{\tau\tau}| \leq m_2|U_{32}|^2+m_3|U_{33}|^2 \ (\text{NO}),\\
\left|m_1|U_{31}|^2-m_2|U_{32}|^2\right| &\leq |M_{\tau\tau}| \leq m_1|U_{31}|^2+m_2|U_{32}|^2 \ (\text{IO}).
\end{align}
By using the best-fit values  (with Super-Kamiokande atmospheric data) in Ref. \cite{Esteban:2020cvm},
one finds
\begin{align}
0.0180\ \text{eV} &\leq |M_{\tau\tau}| \leq 0.0238\ \text{eV} \ (\text{NO}),\\
0.0126\ \text{eV} &\leq |M_{\tau\tau}| \leq 0.0291 \ \text{eV} \ (\text{IO})\;, \label{emnui}
\end{align}
so we use $|M_{\tau\tau}|=0.02$ eV as a benchmark. 

In the IO case, by choosing $\delta\sim \pi, \eta\sim \frac{\pi}{2}$ and tuning the other mixing parameters, $M_{\tau\tau}$ can be much smaller than 0.0126 eV. The following discussion includes such a case. 
Note that even if $M_{\tau\tau}=0$, $m_{S_1}$ satisfies the bounds (\ref{ems1_mutau}) and (\ref{ems1_mumu}) coming from the other components.

\section{Lepton flavor violating decays} \label{sbr}
\subsection{Lower bounds on branching ratios}
In this section, we discuss the LFV constraint on the model.
We focus on the LFV branching ratios of $\ell_i\to \ell_j\gamma$. 
The decay width is given by 
\begin{equation}
  \Gamma(\ell_i\to \ell_j\gamma)=\frac{\alpha_{em}}{4}m_{\ell_i}^5\left(|A_L^{ij}|^2+|A_R^{ij}|^2\right), \label{ega}
\end{equation}
where $A_R^{ij}$ and $A_L^{ij}$ are
\begin{align}
  A_R^{ij}=&\frac{1}{16\pi^2m_{S_2}^2}\sum_{I=1}^{n_N}g_{Ii}^*g_{Ij}F_2(x_I), \label{ear}\\
  A_L^{ij}=&\frac{1}{16\pi^2m_{S_1}^2}\sum_{k=1}^3h_{ik}h_{jk}^*F_2(0)
  =\frac{1}{192\pi^2m_{S_1}^2}h_{il}h_{jl}^*\; ,
\end{align}
where $l\neq i,j$. In the above expression, the loop function $F_2(x)$ is defined 
as~\cite{Bertolini:1990if}\footnote{
  This function $F_2(x)$ differs from the $F_1(x)$ in Ref.~\cite{Chowdhury:2018nhd} by factor 2, i.e., $F_2(x)=\frac{1}{2}F_1(x)$.
} 
\begin{equation}
  F_2(x)= \frac{2x^2+5x-1}{12(x-1)^3}-\frac{x^2\log(x)}{2(x-1)^4}\;.
\end{equation}
By using the decay rate
\begin{equation}
  \Gamma(\ell_i\to \ell_j\nu\bar{\nu})\simeq \frac{G_F^2m_{\ell_i}^5}{192\pi^3},
\end{equation}
we obtain the branching ratio
\begin{align}
\text{Br}(\ell_i\to \ell_j\gamma)&=\frac{\Gamma(\ell_i\to \ell_j\gamma)}{\Gamma(\ell_i\to\ell_j\nu\bar{\nu})}\textrm{Br}(\ell_i\to \ell_j \nu \bar{\nu}) \nonumber \\
&=\frac{48\pi^3\alpha_{em}}{G_F^2}\left(|A_L^{ji}|^2+|A_R^{ij}|^2\right)\text{Br}(\ell_i\to \ell_j\nu\bar{\nu}). \label{ebr0}
\end{align}

First, let us consider the $S_2$ contribution $A_R$. 
In general, there is only a trivial bound $|A_R^{ij}|\geq 0$, and we use it in this section. For example, if we set  $g_{I1}=0$, $A_R^{21}=A_R^{31}=0$ is realized. With this choice of the couplings, however, another column $g_{I2}$ or $g_{I3}$ cannot be zero simultaneously to reproduce the rank two neutrino mass matrix, so $\tau\to \mu\gamma$ can be significant.
We will discuss this point in Sec.~\ref{secBP}. 

Next, we consider the $S_1$ exchanging contribution $A_L$.
As shown in Ref. \cite{Irie:2021obo}, any off-diagonal component $h_{ij}$ cannot be zero to produce neutrino oscillation parameters. Furthermore, $A_L$ is proportional to $m_{S_1}^{-2}$, while $m_{S_1}$ has an upper bound [Eq.~(\ref{ems1_2})].
Therefore, $A_L$ has a nontrivial lower bound. 

By imposing Eq.~(\ref{ems1_2}) and $|A_R^{ij}|\geq 0$ to Eq. (\ref{ebr0}), we find
\begin{align}
\textrm{Br}(\ell_i\to \ell_j \gamma) &>\frac{\alpha_{em}}{768\pi G_F^2}\left(\frac{4(4\pi)^4M_{\tau\tau}}{1.05n_{\text{eff}}m_\mu^2h_{23}^2}\right)^4|h_{il} h_{jl}|^2\textrm{Br}(\ell_i\to \ell_j \nu \bar{\nu}) \nonumber \\
&=  7.45\times 10^{-16} \frac{|h_{il} h_{jl}|^2}{|h_{23}|^8 n_{\text{eff}}^4} \left(\frac{|M_{\tau\tau}|}{\rm 0.02 eV}\right)^4 \textrm{Br}(\ell_i\to \ell_j \nu \bar{\nu}).
\label{ebrij}
\end{align}

The ranges of $h_{ij}$ are determined by neutrino oscillation parameters. 
We define $k$ and $k'$ as 
\begin{align}
k\equiv \frac{h_{12}}{h_{23}}\;,\quad
k'\equiv \frac{h_{13}}{h_{23}} \;,
\label{defk}
\end{align}
and by solving Eq. (\ref{eq:KNTneutrinomassmatrix}), they can be expressed by the neutrino mass matrix components as\footnote{It can be derived straightforwardly using Eqs (10) and (11) in Ref. \cite{Irie:2021obo}.}
\begin{align}
k &= \frac{M_{e\mu}M_{\mu\tau}-M_{e\tau}M_{\mu\mu}}{M_{\mu\mu}M_{\tau\tau}-M_{\mu\tau}^2}, \label{ek}\\
k' &= \frac{M_{e\mu}M_{\tau\tau}-M_{e\tau}M_{\mu\tau}}{M_{\mu\mu}M_{\tau\tau}-M_{\mu\tau}^2}.\label{ekp}
\end{align}
The ranges of these parameters significantly depend on the neutrino mass ordering. 

For the NO case, the $3\sigma$ ranges of $k$ and $k'$ are $0.27\leq |k|\leq 0.67$ and $0.26\leq |k'|\leq 0.66$ \cite{Irie:2021obo}. Since $|k|,|k'|<1$, the perturbativity should be imposed to the largest component: $|h_{23}|<1$. By this condition, the factor of $h_{ij}$ in the branching ratios have to satisfy
\begin{align}
\frac{|h_{23} h_{13}|^2}{|h_{23}|^8 } &=\frac{|k'|^2}{|h_{23}|^4}> |k'|^2, \\
\frac{|h_{23} h_{12}|^2}{|h_{23}|^8 }&=\frac{|k|^2}{|h_{23}|^4} > |k|^2, \\
\frac{|h_{12} h_{13}|^2}{|h_{23}|^8 } &=\frac{|kk'|^2}{|h_{23}|^4} > |kk'|^2.
\end{align}
Finally, using the best-fit values of the experimental data \cite{ParticleDataGroup:2020ssz}
\begin{align}
\text{Br}(\mu\to e\nu\bar{\nu}) &= 1,\\
\text{Br}(\tau\to e\nu\bar{\nu}) &= 0.1782\pm 0.0004,\\
\text{Br}(\tau\to \mu\nu\bar{\nu}) &= 0.1739\pm 0.0004,
\end{align}
we obtain
\begin{align}
\textrm{Br}(\mu\to e \gamma) &> 
5.0\times 10^{-18}\left(\frac{|M_{\tau\tau}|}{\rm 0.02 eV}\right)^4\left(\frac{n_{\text{eff}}}{2}\right)^{-4}\left(\frac{|k'|}{0.329}\right)^2, \\
\textrm{Br}(\tau\to e \gamma) &> 
3.0\times 10^{-18}\left(\frac{|M_{\tau\tau}|}{\rm 0.02 eV}\right)^4\left(\frac{n_{\text{eff}}}{2}\right)^{-4}\left(\frac{|k|}{0.600}\right)^2,\\
\textrm{Br}(\tau\to \mu \gamma) &> 
3.2\times 10^{-19}\left(\frac{|M_{\tau\tau}|}{\rm 0.02 eV}\right)^4\left(\frac{n_{\text{eff}}}{2}\right)^{-4}\left(\frac{|k|}{0.600}\right)^2\left(\frac{|k'|}{0.329}\right)^2,
\end{align}
where $k$ and $k'$ are factored out by using best-fit values. 

For the IO case, the $3\sigma$ ranges of $k$ and $k'$ are $3.9\leq |k|\leq 5.3$ and $4.0\leq |k'|\leq 5.4$ \cite{Irie:2021obo}.
Using the best-fit parameters, we obtain 
$|k|= 4.31$ and $|k'|=5.01$. Since $|k'|>|k|,1$, we impose the perturbativity condition as $|h_{13}|<1$. By this condition, we obtain
\begin{align}
\frac{|h_{23} h_{13}|^2}{|h_{23}|^8 } &=\frac{|k'|^6}{|h_{13}|^4} >|k'|^6,\\
\frac{|h_{23} h_{12}|^2}{|h_{23}|^8 } &= \frac{|k|^2|k'|^4}{|h_{13}|^4}> |k|^2|k'|^4, \\
\frac{|h_{12} h_{13}|^2}{|h_{23}|^8 } &=\frac{|k|^2|k'|^6}{|h_{13}|^4}> |k|^2|k'|^6.
\end{align}
By substituting these inequalities to Eq.~(\ref{ebrij}), the lower bounds are
\begin{align}
\textrm{Br}(\mu\to e \gamma) &> 
7.4\times 10^{-13}\left(\frac{|M_{\tau\tau}|}{\rm 0.02 eV}\right)^4\left(\frac{n_{\text{eff}}}{2}\right)^{-4}\left(\frac{|k'|}{5.01}\right)^6, \label{ebrmu}\\
\textrm{Br}(\tau\to e \gamma) &> 
9.7\times 10^{-14}\left(\frac{|M_{\tau\tau}|}{\rm 0.02 eV}\right)^4\left(\frac{n_{\text{eff}}}{2}\right)^{-4}\left(\frac{|k|}{4.31}\right)^2\left(\frac{|k'|}{5.01}\right)^4,\\
\textrm{Br}(\tau\to \mu \gamma) &> 
2.4\times 10^{-12}\left(\frac{|M_{\tau\tau}|}{\rm 0.02 eV}\right)^4\left(\frac{n_{\text{eff}}}{2}\right)^{-4}\left(\frac{|k|}{4.31}\right)^2\left(\frac{|k'|}{5.01}\right)^6.
\end{align}
In the case with three right-handed neutrinos
$n_{\text{eff}}\leq 2$ and the best-fit neutrino oscillation parameters $|M_{\tau\tau}|\geq 0.126$~eV [see Eq. (\ref{emnui})], the bounds are
\begin{align}
\textrm{Br}(\mu\to e \gamma) &> 
1.1\times 10^{-13}, \label{ebrbest}\\
\textrm{Br}(\tau\to e \gamma) &> 
1.5\times 10^{-14},\\
\textrm{Br}(\tau\to \mu \gamma) &> 
3.7\times 10^{-13}.
\end{align}

As a whole, the bounds are more severe for the IO case than the NO case. It comes from the fact that $h_{23}$ 
is smaller for the IO case to produce the neutrino mass matrix. Even the bounds for the NO case, however, are 
much stronger than the contribution of active neutrinos ${\rm Br}(\mu\to e\gamma)<10^{-54}$ \cite{Petcov:1976ff,Marciano:1977wx,Lee:1977qz,Lee:1977tib,deGouvea:2013zba}.

\subsection{Constraints on the parameters}

\begin{figure}[htbp]
    \begin{tabular}{cc}
      \begin{minipage}[t]{0.45\hsize}
        \centering
        \includegraphics[clip, width=70mm]{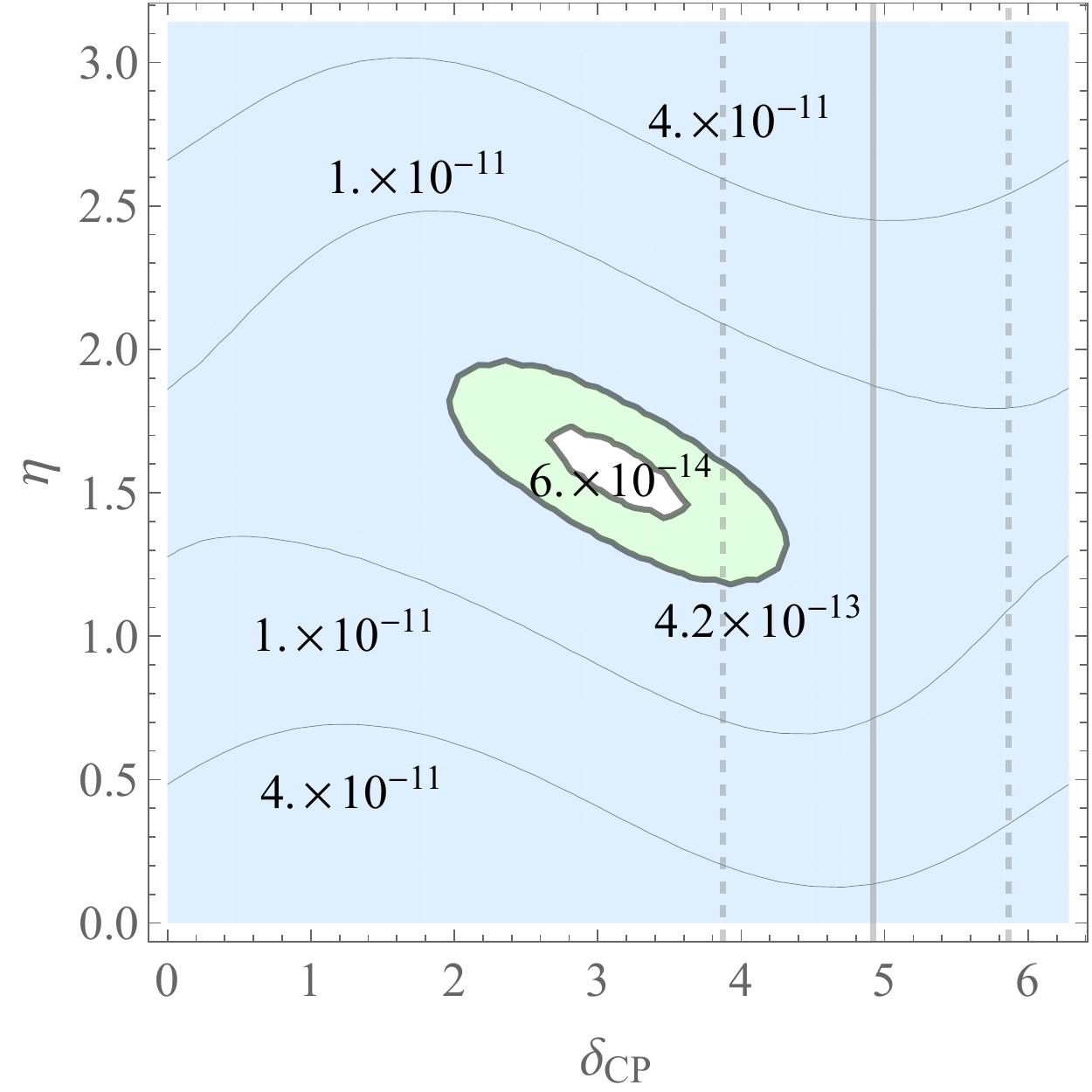}
        \subcaption{$n_\text{eff}=1$, best-fit}
        \label{fn1bfeta}
      \end{minipage} &
      \begin{minipage}[t]{0.45\hsize}
        \centering
        \includegraphics[clip, width=70mm]{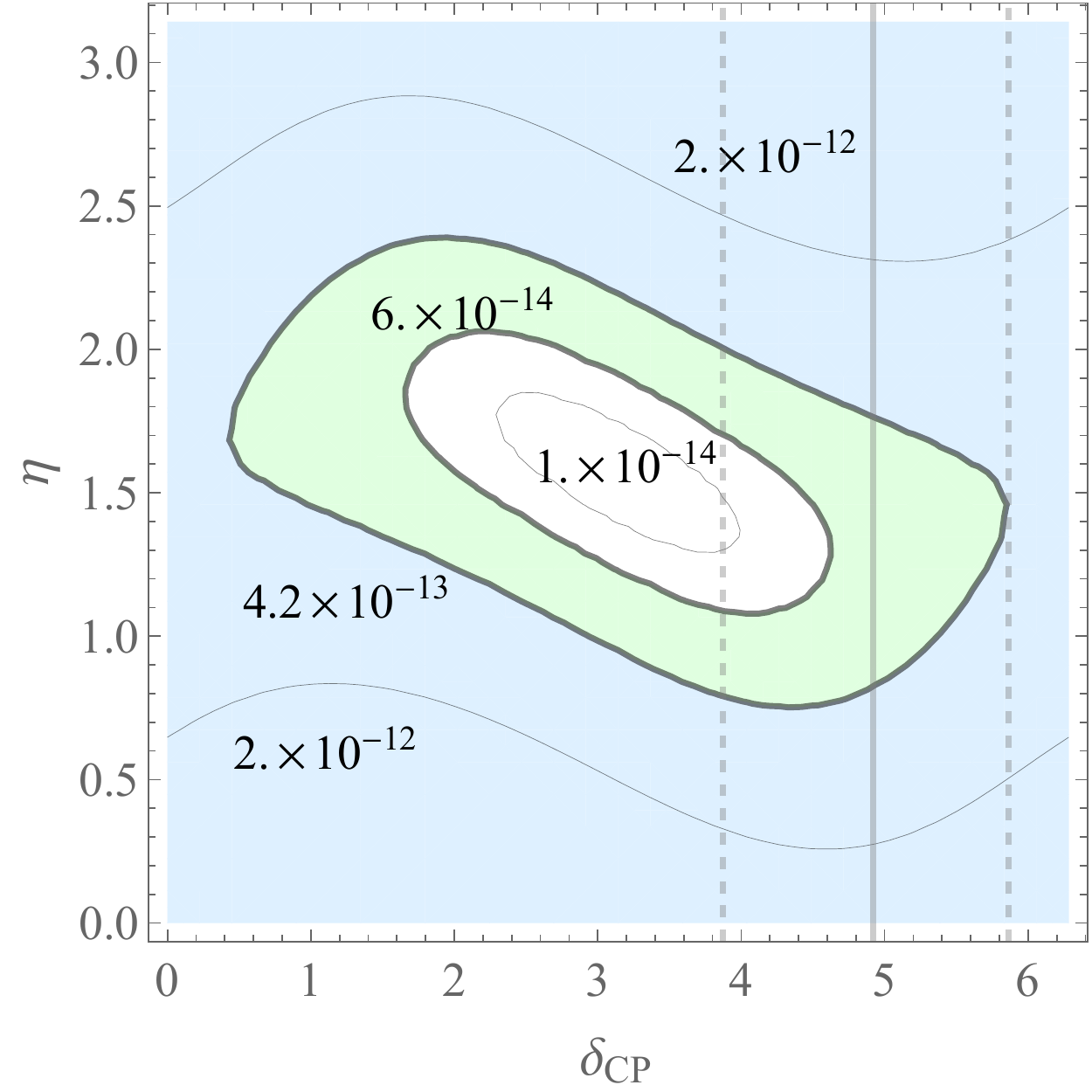}
        \subcaption{$n_\text{eff}=2$, best-fit}
        \label{fn2bfeta}
      \end{minipage} \\
   \\
      \begin{minipage}[t]{0.45\hsize}
        \centering
        \includegraphics[clip, width=70mm]{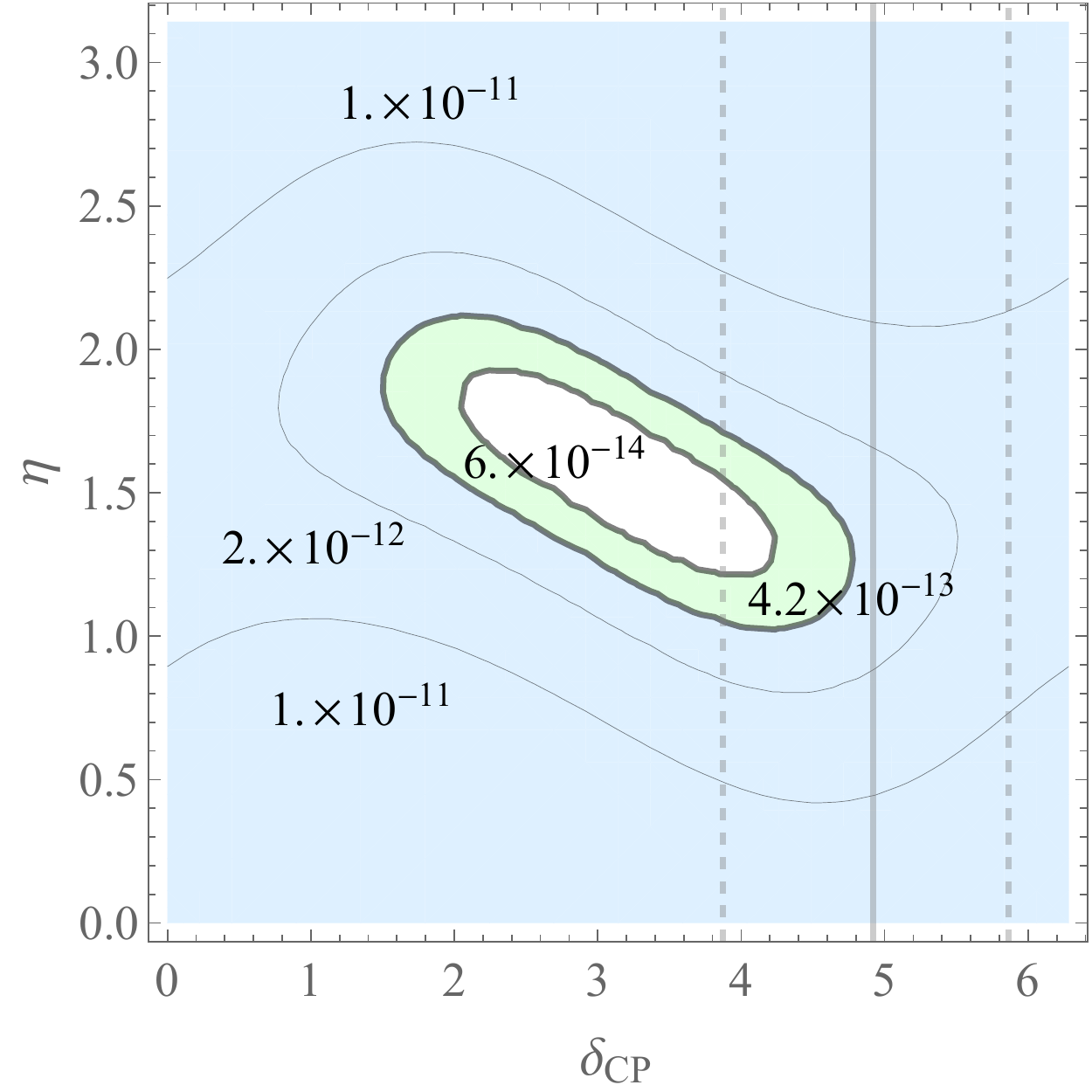}
        \subcaption{$n_\text{eff}=1$, minimal}
        \label{fn1mineta}
      \end{minipage} &
      \begin{minipage}[t]{0.45\hsize}
        \centering
        \includegraphics[clip, width=70mm]{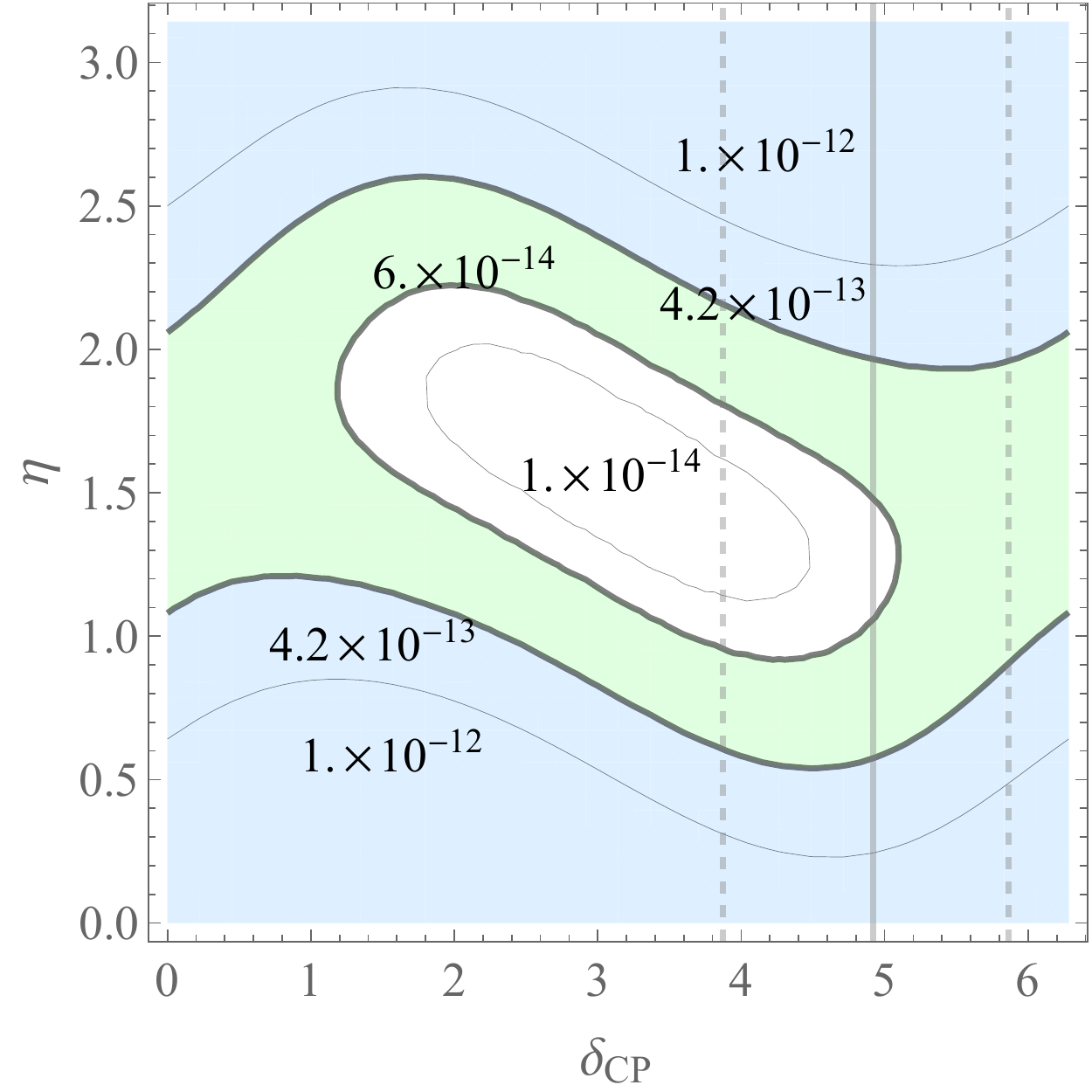}
        \subcaption{$n_\text{eff}=2$, minimal}
        \label{fn2mineta}
      \end{minipage} 
    \end{tabular}
     \caption{Contour plots of the lower bounds on Br($\mu\to e\gamma$) in the inverted mass ordering cases. $n_\text{eff}\ (\leq n_N-1)$ is the number of right-handed neutrinos that contribute to $M_{\tau\tau}$. In cases (a) and (b),  the other neutrino oscillation parameters are fixed to the best-fit values \cite{Esteban:2020cvm}. In cases (c) and (d), they are chosen to realize minimal Br($\mu\to e\gamma$) within two standard deviations. The blue regions are already excluded by MEG \cite{MEG:2016leq}. The green regions can be excluded by MEG II \cite{MEGII:2021fah}. The solid- and dashed-vertical lines indicate the best-fit value and $2\sigma$ ranges of $\delta$.}
     \label{fetad}
  \end{figure}

\begin{figure}[htbp]
    \begin{tabular}{cc}
      \begin{minipage}[t]{0.45\hsize}
        \centering
        \includegraphics[clip, width=70mm]{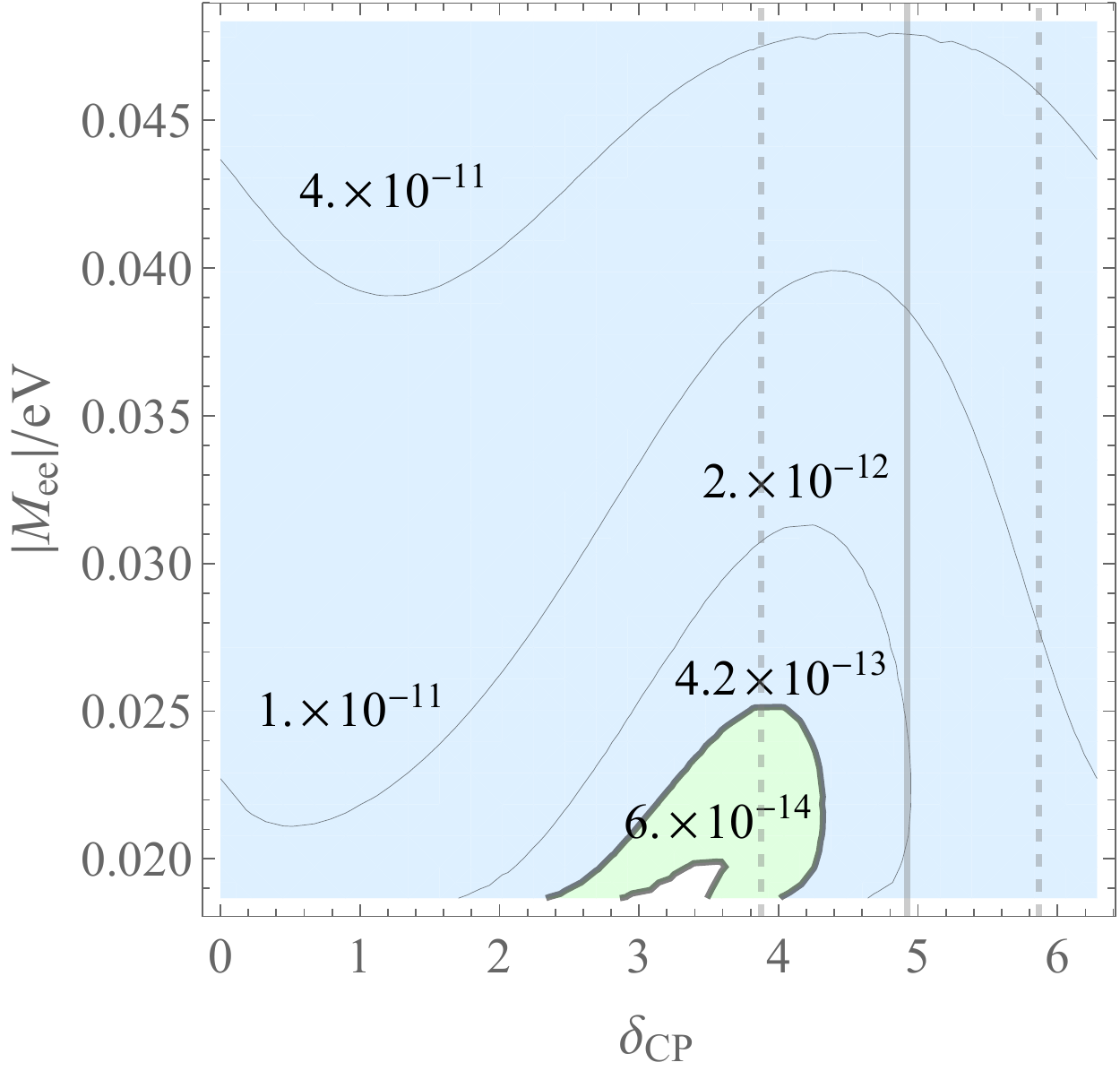}
        \subcaption{$n_\text{eff}=1$, best-fit}
        \label{fn1bf}
      \end{minipage} &
      \begin{minipage}[t]{0.45\hsize}
        \centering
        \includegraphics[clip, width=70mm]{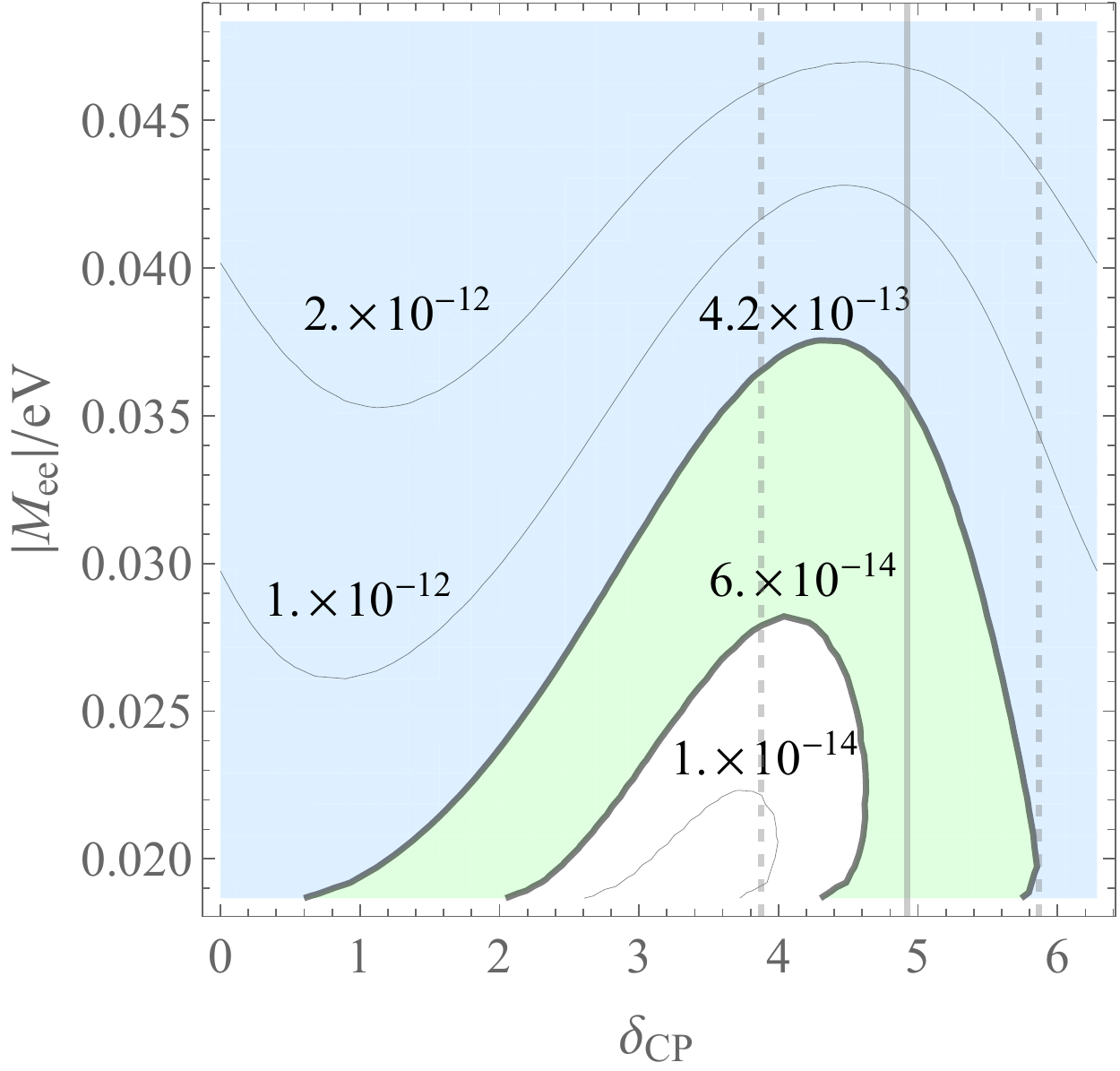}
        \subcaption{$n_\text{eff}=2$, best-fit}
        \label{fn2bf}
      \end{minipage} \\
   \\
      \begin{minipage}[t]{0.45\hsize}
        \centering
        \includegraphics[clip, width=70mm]{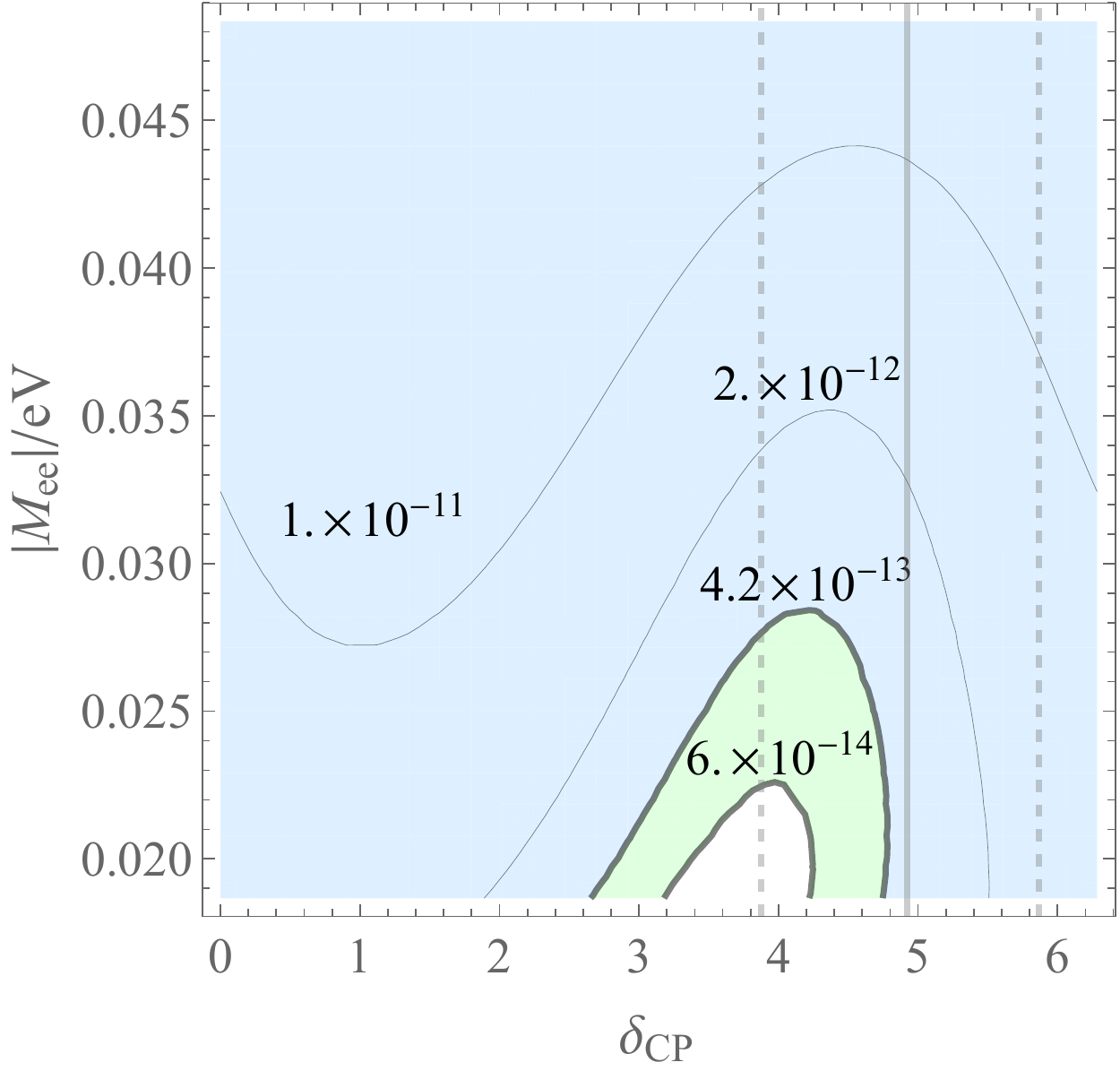}
        \subcaption{$n_\text{eff}=1$, minimal}
        \label{fn1min}
      \end{minipage} &
      \begin{minipage}[t]{0.45\hsize}
        \centering
        \includegraphics[clip, width=70mm]{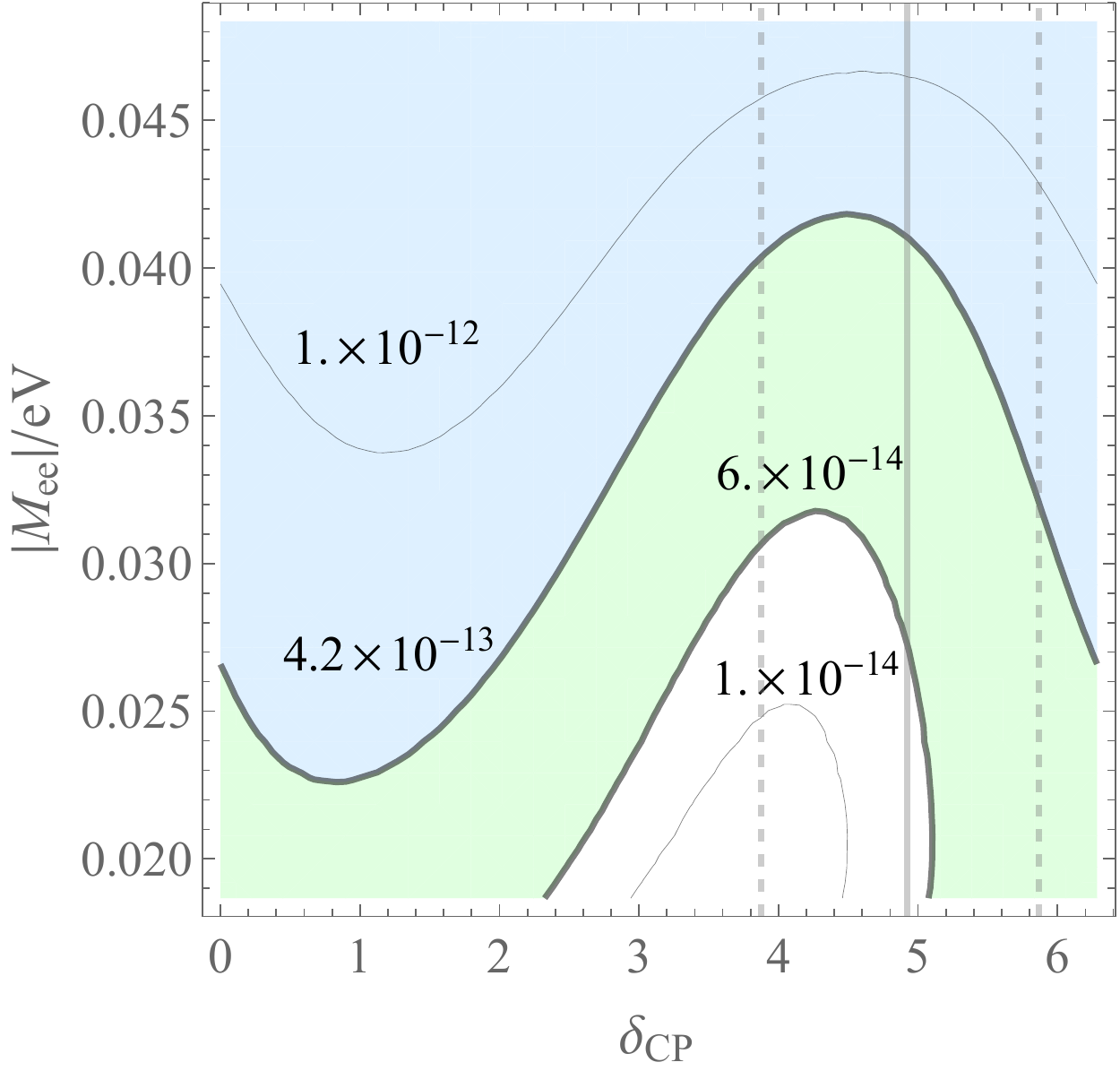}
        \subcaption{$n_\text{eff}=2$, minimal}
        \label{fn2min}
      \end{minipage} 
    \end{tabular}
     \caption{Contour plots of the lower bound on Br($\mu\to e\gamma$) in the IO case. Notation is same as Fig.~\ref{fetad}. The Majorana phase $\eta$ is converted to $|M_{ee}|$ by Eq. (\ref{ecos2eta}).}
     \label{fmeed}
  \end{figure}

Experimental upper limits of the LFV branching ratios (90\% confidence level) are \cite{ParticleDataGroup:2020ssz,MEG:2016leq,BaBar:2009hkt}
\begin{align}
\textrm{Br}(\mu\to e \gamma) &< 4.2\times 10^{-13}, \label{eMEG}\\
\textrm{Br}(\tau\to e \gamma) &< 3.3\times 10^{-8},\\
\textrm{Br}(\tau\to \mu \gamma) &< 4.4\times 10^{-8}.
\end{align}
These lepton flavor violations will be explored by many future experiments. The sensitivity of MEG II on this $\mu \rightarrow e \gamma$ mode is expected to be $6\times 10^{-14}$ \cite{MEGII:2021fah}. The sensitivity of Belle-II on the LFV mode with $\tau$ is estimated to $O(10^{-9})$ - $O(10^{-10})$ \cite{Belle-II:2018jsg}. In this subsection, we focus on $\mu \rightarrow e \gamma$ and consider the IO case.

To see the behavior of the lower bound, we write $|M_{\tau\tau}|$ explicitly:
\begin{align}
|M_{\tau\tau}|= \left|m_1 (s_{12}s_{23}-c_{12}s_{13}c_{23}e^{i\delta})^2e^{2i\eta}+m_2(c_{12}s_{23}+s_{12}s_{13}c_{23}e^{i\delta})^2\right|. \label{emtt2}
\end{align}
It largely depends on $\delta$ and $\eta$. For example, if $\delta\sim\pi$ and $\eta\sim \pi/2$, a cancellation can happen. 

In Fig.~\ref{fetad}, we plot the lower bound in the $\delta-\eta$ planes using the bound~(\ref{ebrmu}) and Eq. (\ref{emtt2}). Figs.~\ref{fn1bfeta} and~\ref{fn2bfeta} are the cases that oscillation parameters are set to the best-fit values. Most regions of the parameters are already excluded by the constraint from the MEG experiment (blue region). 
In particular, if $\delta$ is also the best-fit value, $n_\text{eff}=1$ is excluded, and $n_\text{eff}=2$ can be excluded by MEG II. As we stated, the region including $\delta=\pi$ and $\eta=\pi/2$ is unconstrained by the LFV experiments.

We also searched for the neutrino parameters that minimizes the right-hand side of the inequality~(\ref{ebrmu}) within two standard deviation ranges.\footnote{To obtain the ranges, we simply multiplied the standard deviation in Ref.~\cite{Esteban:2020cvm} by two. We chose two standard deviations because the constraints on Br($\mu\to e\gamma$) are given at 90 \% C.L.} We found that parameters 
\begin{align}
\begin{aligned}
\Delta m_{21}^2 &= 7.03\times 10^{-5} \text{eV}^2,\ \Delta m_{32}^2 = -2.442\times 10^{-3}\text{eV}^2, \\
\theta_{12} &= 35.01^\circ,\ \theta_{23} = 47.1^\circ,\  \theta_{13} = 8.84^\circ,\\
\delta &= 222^\circ,\ \eta = 79.1^\circ
\end{aligned}
\end{align}
realize the minimum:
\begin{align}
\textrm{Br}(\mu\to e \gamma) &> 
1.0\times 10^{-14} n_\text{eff}^{-4}.
\end{align}
Figures~\ref{fn1mineta} and~\ref{fn2mineta} show the lower bound with these parameters (except $\delta$ and $\eta$). The unconstrained regions are enlarged, but the MEG II experiment can probe most regions even in these conservative cases.

The Majorana phase is searched by neutrinoless double-beta decay experiments, and they are sensitive to the $ee$ component of the neutrino mass matrix:
\begin{align}
M_{ee}= (m_1 c_{12}^2 e^{2i\eta}+m_2 s_{12}^2)c_{13}^2.
\end{align}
This is independent from $\delta$, and determined by $\eta$. We can express $\eta$ by $|M_{ee}|$ as
\begin{align}
\cos 2\eta=\frac{1}{2m_1 m_2 s_{12}^2c_{12}^2}\left(\frac{|M_{ee}|^2}{c_{13}^4}-m_1^2 c_{12}^4-m_2 c_{12}^4\right). \label{ecos2eta}
\end{align}
The range of $|M_{ee}|$ which satisfies $|\cos2\eta|\leq 1$ is (using the best-fit values)
\begin{align}
0.0187\ \text{eV}\leq|M_{ee}|\leq 0.0484 \ \text{eV}. \label{emeerange}
\end{align}
We plot the lower bound of Br($\mu\to e\gamma$) as a function of $|M_{ee}|$ and $\delta$ in Fig. \ref{fmeed}. 

The present upper limit on $|M_{ee}|$ by neutrinoless double beta decay experiments is 66 to 155 meV \cite{GERDA:2019ivs}, so they do not constrain the range given by Eq.~(\ref{emeerange}). 
The sensitivities of future experiments (90\% CL) are, however, 19 meV-46 meV (SNO+ Phase II \cite{SNO:2015wyx}), 5.7 meV-17.7 meV (nEXO, after 10 years of data taking \cite{nEXO:2017nam}), so the all range of $|M_{ee}|$ can be searched. 
These experiments, in combination with neutrino oscillation and LFV experiments, can exclude the KNT model with $n_N=3$ even if $\delta$ is not the best-fit value today. For instance, if neutrinoless double-beta decay experiments find $|M_{ee}|=45$ meV and neutrino oscillation experiments show the mass ordering is inverted, the KNT model with $n_N=3$ predicts too large Br($\mu\to e\gamma$) and the model is excluded.

\section{Dark matter property and LFV}
\label{secBP}

In this section, we discuss a case that Br($\mu\to e\gamma$) is close to the lower bound given in the previous section to avoid severe experimental constraints. As stated after Eq.~(\ref{ebr0}), the $A_R^{32}$ term in the Br($\tau\to \mu\gamma$) can be large in this situation. The loop function $F_2$ in $A_R$ is monotonically decreasing, so the lightest right-handed neutrino $N_1$ contributes to $\tau\to\mu\gamma$ dominantly. We first discuss a constraint on $N_1$ as dark matter, then show minimal Br($\tau\to \mu\gamma$) with explicit parameters in the allowed region.

\subsection{Dark matter bound}\label{secdm}
In the KNT model, the lightest $Z_2$ odd particle is stable.
When the $Z_2$ odd particle mass spectrum satisfies $m_{N_1}<m_{S_2}$, $N_1$ is the dark matter candidate. 
The thermal relic abundance of $N_1$ in the early Universe is constrained 
as $\Omega_{N_1}h^2\lesssim 0.1$.
The abundance approximately depends on the 
annihilation cross section $\sigma v$ as 
\begin{align}
  \Omega_{N_1}h^2\sim 0.1\left(\frac{3\times 10^{-26}~\text{cm}^3/\text{s}}{\sigma v}\right)\;,
\end{align}
so $\sigma v\gtrsim 3\times 10^{-26}~\text{cm}^3/\text{s}$ is needed.

In the most region of the parameter space, the dominant annihilation mode is 
$N_1 N_1 \to \ell_i\bar{\ell_j}$ via $t$- and $u$-channel exchanges of $S_2$. The cross section is given by 
\begin{align}
  \sigma v \simeq \frac{m_{N_1}^2(m_{N_1}^4+m_{S_2}^4)}{8\pi (m_{N_1}^2+m_{S_2}^2)^4}x_f
  \sum_{i=1}^3\sum_{j=1}^{3}|g_{1i}^*g_{1j}|^2
  \;,
\end{align}
where $x_f\equiv T_f/m_{N_1}$ is determined by the freeze-out temperature of $N_1$, $T_f$, and $x_f\sim 1/20$.
When $m_{N_1}\simeq m_{S_2}$ is satisfied, the coannihilation process with $S_2$ 
gives a significant contribution. 

In Fig.~\ref{fig:TMG}, we show the contours of $\sigma v/\sum_i\sum_j|g_{1i}^*g_{1j}|^2$ in the $m_{N_1}$-$m_{S_2}$ plane. 
For $g_{11}=0$, the dark matter abundance condition and the perturbativity $|g_{12}|<1,\ |g_{13}|<1$ give $m_{N_1}\leq m_{S_2}\lesssim 750$~GeV.

\subsection{Benchmark example}

\begin{figure}
  \begin{center}
  \includegraphics[width=8cm]{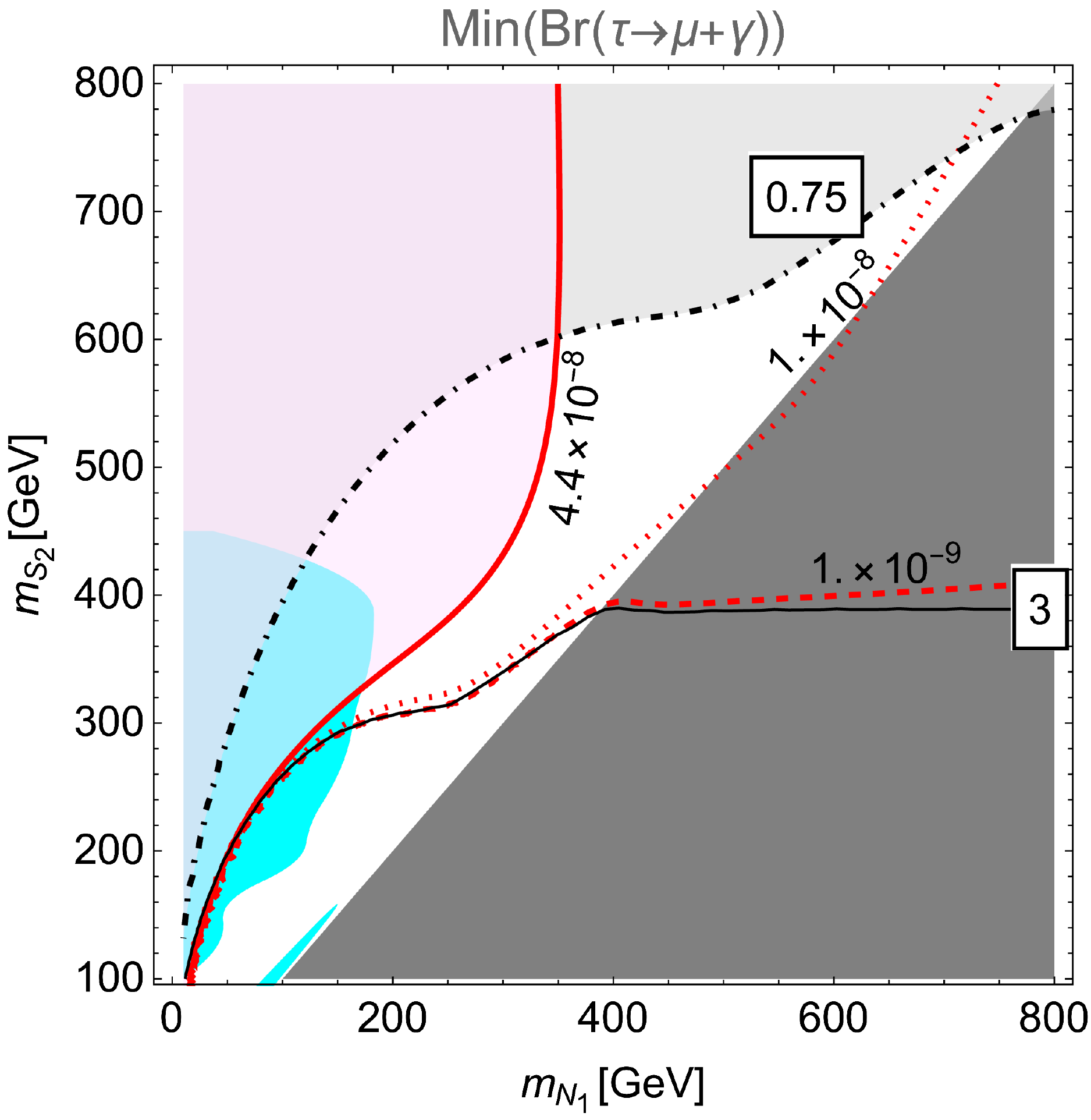}
  \end{center}
  \caption{Contour of the minimal value of $\text{Br}(\tau\to \mu\gamma)$ (red lines). The light-red region is already excluded by the BABAR experiment \cite{BaBar:2009hkt}.
  The dark gray region shows where the charged particle $S_2$
  is stable. The lines labeled ``0.75'' and ``3'' indicate $\sigma v/(\sum |g_{1i}^*g_{1j}|^2\ (10^{-26} \text{cm}^3/\text{s}))$, and
  the light-gray region upper of the dot-dashed curve, 
  either $|g_{12}|>1$ or $|g_{13}|>1$ is required to satisfy $\Omega_{N_1} h^2\simeq 0.1$.
  The cyan region is excluded by the direct search of the right-handed slepton 
  at the LHC\cite{ATLAS:2019lff,ATLAS:2019lng}. 
  }
  \label{fig:TMG}
\end{figure}

Let us define our benchmark point.
To take Br($\mu\to e \gamma$) as small as possible, we need large loop functions $f(x_{2,3},y)$, i.e., $m_{N_{2,3}}=5.43m_{S_1}\gg m_{S_2}$, and large couplings. We choose 
\begin{align}
m_{S_1} &= 8.74\times 10^{4}\ \text{GeV},\quad
M_{N_2}=M_{N_3} = 4.75\times 10^5\ \text{GeV},\quad
\lambda_S =\max(|h_{ij}|)=1\;,
\end{align}
and parametrize the Yukawa matrix $g_{Ij}\ (n_N=3)$ as 
\begin{equation}
  g=\begin{pmatrix}
    0&g_{12}&g_{13}\\
    0&1&g_{23}\\
    0&1&g_{33}
  \end{pmatrix}\;,
\end{equation}
where we set $g_{I1}=0$ to suppress the $S_2$ exchange contribution to $\mu\to e\gamma$.
Once we input the neutrino masses, the parameters in the mixing matrix, and $m_{S_2}$, 
the parameters $h_{ij}$, $g_{23}$, $g_{33}$  are determined by the procedure given in Ref.~\cite{Irie:2021obo}.
On the other hand, $m_{N_1}$, $g_{12}$, $g_{13}$ are almost irrelevant to the neutrino mass matrix, since the $N_1$ contribution to the neutrino mass matrix is significantly suppressed due to 
$f(x_1,y)\ll f(x_{2},y)=f(x_3,y)$.
The parameters $g_{23}$ and $g_{33}$ are related to the neutrino mass components as 
\begin{align}
  -\frac{M_{\mu\mu}}{M_{\mu\tau}}=\frac{m_{\tau}}{m_{\mu}}\frac{X_{33}}{X_{23}}\;,\quad 
  -\frac{M_{\tau\tau}}{M_{\mu\tau}}=\frac{m_{\mu}}{m_{\tau}}\frac{X_{22}}{X_{23}}\;,
  \label{eqMnuX}
\end{align}
with 
\begin{align}
  X_{22}=&\ \sum_{I=1}^3g_{I2}^2f(x_I,y)
  \simeq 2f(x_2,y)\;,\quad \nonumber\\
  X_{23}=&\ \sum_{I=1}^3g_{I2}g_{I3}f(x_I,y)
  \simeq (g_{23}+g_{33})f(x_2,y)\;,\quad\nonumber\\
  X_{33}=&\ \sum_{I=1}^3g_{I3}^2f(x_I,y)
  \simeq (g_{23}^2+g_{33}^2)f(x_2,y)\;.
\end{align}
By solving Eq.~(\ref{eqMnuX}), $g_{23}$ and $g_{33}$ are determined.
Taking into account $\max(|h_{ij}|)=1$, the matrix $h$ is also determined by 
Eqs.~(\ref{defk}), (\ref{ek}), and (\ref{ekp}).

For example, by using the best-fit neutrino parameters in Ref.~\cite{Esteban:2020cvm}, $\eta = 1.29$, and $m_{S_2}=300$~GeV, 
we obtain
\begin{align}
g &=
\begin{pmatrix}
0 & g_{12} & g_{13}\\
0 & 1 & (1.2+2.8i)\times 10^{-2} \\
0 & 1 & (8.1+3.5i)\times 10^{-2}
\end{pmatrix}, \label{egmat}
\end{align}
and 
\begin{align}
  h = 
\begin{pmatrix}
0&-0.860 &1\\
0.860 & 0 & -0.041+0.195i\\
-1 & 0.041 - 0.195i & 0
\end{pmatrix}.
\end{align}
In this case, 
the branching ratio and the effective neutrino mass are calculated as
\begin{align}
\text{Br}(\mu\to e\gamma) &=1.2\times 10^{-13}, \\
|M_{ee}| &= 0.0224 \text{ eV}.
\end{align}
These are below the constraints today but can be tested in future experiments.

In our benchmark case, the $S_1$ contribution to the LFV is suppressed enough
to satisfy the current experimental bound.
Hereafter, we focus on the $S_2$ contribution to the LFV.
There is no contribution to $\mu\to e\gamma$ and $\tau\to e\gamma$ because of 
our ansatz $g_{I1}=0$, 
while contribution to $\tau\to \mu \gamma$ can be significant. In Fig.~\ref{fig:TMG}, we display the contour of the minimal value of 
$\text{Br}(\tau\to \mu\gamma)$ for each parameter point of $(m_{N_1}, m_{S_2})$.
We have taken into account our ansatz that all the dimensionless coupling constants 
should be less than one.
For larger $m_{N_1}$ and $m_{S_2}$, $|g_{12}|^2+|g_{13}|^2$ 
needs to be larger in order to reproduce the dark matter relic abundance.
In the upper region of the dot-dashed curve labeled ``1.5'', $|g_{12}|>1$ or $|g_{13}|>1$ 
is required.
The left upper region of the solid red curve is 
already excluded by the BABAR experiment \cite{BaBar:2009hkt}. 
In near future, the experimental sensitivity for $\text{Br}(\tau\to \mu\gamma)$ 
is expected to be improved factor 100 at the Belle II experiment \cite{Belle-II:2018jsg}, 
and the wide region of the 
parameter space can be explored.

\section{Conclusions} \label{scon}
In this paper, we have discussed the constraints on the parameter space of the KNT model by taking into account the perturbativity 
of the dimensionless coupling constants and neutrino mass matrix components. 
We have found that there are lower limits of the predicted $\text{Br}(\mu\to e\gamma)$, $\text{Br}(\tau\to e\gamma)$, and $\text{Br}(\tau\to \mu\gamma)$ which are induced by the $S_1$ scalar exchange.
In the IO case, the bound on Br($\mu\to e\gamma$) is so severe that the
wide ranges of the Dirac and Majorana CP phases are restricted. If the neutrino oscillation parameters are best-fit values in the IO case, $n_\text{eff}=1$ is already excluded and $n_\text{eff}=2$ can be excluded by MEG II.

We have also considered the $S_2$ contribution to $\tau\to \mu\gamma$ in a case with suppressed $\mu\to e\gamma$. 
We have shown that Br($\tau\to \mu\gamma$) can be large enough that wide parameter space can be tested by future experiments such as the Belle II experiment. If $m_{N_1}, m_{S_2}<350$~GeV, Br($\tau\to \mu\gamma$) can be suppressed, but such light particles can be directly searched at the LHC experiments or 
future $e^+e^-$ collider experiments.

\section*{Acknowledgments}
This work is supported in part by the Japan Society for the Promotion of Science (JSPS) KAKENHI Grants No.~20H00160 (T.S.), No.~19K03860, No.~19K03865 and No.~21H00060 (O.S.).

\end{document}